**RESEARCH ARTICLE**

# COMMUNICATION, AWARENESS AND ACCEPTANCE OF DIGITAL BANKING AMIDST CASH CRUNCH IN SOUTHEAST AND SOUTH-SOUTH, NIGERIA

[1] **Okechukwu Christopher ONUEGBU,** [2] **Bettina Oboakore AGBAMU,** [3] **Belinda Uju ANYAKOHA,** [4] **Ogonna Wilson ANUNIKE**

[1&3] Department of Mass Communication, University of Nigeria, Nsukka. [2]Department of Mass Communication, University of Delta, Agbor. [4]Department of Mass Communication, Federal Polytechnic, Oko, Anambra State, Nigeria

**ABSTRACT**

Digital banking is among the technological innovations currently reverberating the cyber wave. this study seeks to assess communication, awareness and acceptance of it among the residents of south-east and south-south, nigeria. the survey objectives were to ascertain awareness level of the south-east and south-south residents towards digital banking during the cash crunch, determine the acceptance level of digital banking among the south-east and south-south residents, find out the role of communication in awareness and acceptance of digital banking during the cash crunch in south-east and south-south nigeria, and assess the usage of digital banking amidst cash crunch in south-east and south nigeria. the study methodology is a sample survey which allowed researchers to administer questionnaires on 385 respondents out of the 50,166,807 study population. the findings showed that awareness level of digital banking was good (36%) in south-east and south-south nigeria during the cash crunch but it level of acceptance and usage improved more (37%) after the cash crunch. the study also ascertained that communication contribute significantly (59%) towards the usage and acceptance of digital banking in the two zones. it further showed that usage of digital banking in south-east and south-south has improved due to significant contributions of communication.

**Keywords**: cash crunch, communication, digital banking, south-east and south-south Nigeria

**Corresponding Author**
Okechukwu Christopher ONUEGBU
Email Address: oonuegbu3@gmail.com

## 1.0. INTRODUCTION

Digital banking is fast dominating the world. This technologically driven banking service is gaining acceptance from people across drives because it makes it easier for all customers to engage in all kinds of financial and banking transactions at their comfort zones, without having to visit physical banking halls. This includes account opening, monetary deposits, funds transfers, withdrawal, instant loans, checking account balances and amongst other services (Kitsios et al., 2021). In Nigeria, all the commercial banks are currently operating digital banking one way or the other (Salawu et al., 2021). But there are also over 200 fintech and neobanks firms specialising on it in the country thereby making Nigeria a leading or dominating nation in digital banking in the whole of Africa (Guardian, 2023). Among them are Kuda, Opay, Mint, Alat, Sparkle, Vbank, Eyowo, PalmPay, Carbon and Moniepoint.

One striking advantage of digital banks is that they neither have physical branch offices nor staff yet they are accessible, acceptable and available to customers at all time (Kazim, 2023). However, some argued that they are prone to scammers, not readily available to attend to customers' urgent needs especially at night, and poorly used by customers facing power supply challenges unlike the conventional commercial banks like UBA, First Bank and others (Napoletano & Foreman, 2021). Nevertheless, digital banks became heaven or saviour to many Nigerians amidst a cash crunch that made it difficult for people to access physical currency (Joseph, 2023). The awareness and acceptance of digital currency have grown as people seek alternative means of conducting financial transactions, thereby pitching interest with digital banks (Lottu et al., 2023). According to a report by the Nigeria Bureau of Statistics (2021), the total value of e-payment transactions in Nigeria increased by 82% between Q1 2020 and Q1 2021, reaching a total value of N70.2 trillion (approximately $171 billion USD). This growth in e-payment transactions is largely driven by communication, usually through the new media which makes information available to a larger number of people.

The new media has become an empire where people and institutions operate, transact all kinds of businesses and services, while interacting with each other, sharing ideas and information like banking services. In America, for instance, up to 88% of the banks actively utilises social media platforms daily (American Bankers Association, 2023). Also, 15.3 million Nigerians engaged in over 370-million-naira mobile money transactions in 2019 (Adeniyi, 2021). Banks are using it to communicate their new and existing services, respond to customers' queries, increase accessibility, reach out to more people, perform financial transactions, renew and build their reputations (Das, 2024). The fintech and neobanks are using it mostly to acquire new customers, dispense loans and other banking transactions because they are largely active online (on the Internet) (Kazim, 2024).

Although the awareness, acceptance, and usage of digital currency in Nigeria are on the rise, particularly as people seek alternative means of conducting financial transactions in the face of a cash crunch, we cannot situate the place of communication in the trends or specifically state if the rise has affected all parts of Nigeria. While there are still challenges to be addressed, such as the lack of regulatory clarity and the potential for fraud and scams, the potential benefits of digital currency are clear, and it is likely that its use will continue to grow in Nigeria and other countries around the world (Wezel & Ree, 2023), especially because with redesign of new naira notes and the distribution of limited cash and the withdrawal of old currency in circulation, Nigeria has been experiencing a cash crunch in recent times, which has led to a shortage of physical currency.

However, the level of communication, awareness and acceptance of digital banking among the Nigerian population especially among the South East and the South-South residence seems to be relatively low due to lack of data. There seem to be lack of awareness and acceptance that could hinder the growth and adoption of digital banking, which could limit their potential to provide a viable alternative to physical banking in the country.

Therefore, the problem statement is to investigate the place of communication, awareness and acceptance of digital banking among Nigerians in South East, South-South Nigeria, particularly in the context of the cash crunch. The study would be anchored on four objectives such as to ascertain awareness level of the South-East and South-South Nigeria residents towards digital banking during the cash crunch, determine the acceptance level of digital banking among the South-East and South-South Nigeria residents, find out the role of communication in awareness and acceptance of digital banking during the cash crunch in South-East and South-South Nigeria and assess the usage of digital banking amidst cash crunch in South East and South Nigeria. The study questions are what is the level of awareness of the South-East and South-South Nigeria residents of digital banking during the recent cash crunch, the level of acceptance of digital banking among residents of the South East and South-South regions of Nigeria, the role of Communication towards the level of awareness and acceptance of digital banking among the residents of South-East and South-South Nigeria during the cash crunch, and extent digital banking is being used as an alternative means of payment amidst the cash crunch in the South-East and South-South regions of Nigeria.

## 2.0. MATERIALS AND METHODS

The research design employed by this study is a sample survey. This was chosen because it allows researchers to choose and study only a part or sample of the population with a true representation of the whole population. The South East and South-South geopolitical zones or regions of Nigeria were chosen as the area of the study. Located in the Southern part of Nigeria, they were chosen because of diversities in culture, beliefs, tradition, ethnicity, religion and educational backgrounds of the inhabitants.

The two zones comprise eleven States such as Abia, Akwa-Ibom, Anambra, Bayelsa, Cross-River, Delta, Ebonyi, Edo, Enugu, Imo and Rivers. Some languages spoken in the region are Annang, Bini, Effik, Igbo, Ijaw, Ogbia and Urhobo. There are also African Traditional religionists, Christians, Muslims and others dwelling in both regions. Similarly, there are traders, farmers, entrepreneurs, students, public servants, skilled and unskilled manpower in the area due to the presence of various mineral resources and institutions.

The population of the study is 50,166,807. This was extracted from States by State population projection published in 2020 by the National Bureau of Statistics (NBS). A breakdown of the population reports is; Rivers (7,034,973), Anambra (5,599,910), Delta (5,307,543), Imo (5,167,722), Akwa-Ibom (4,780,581), Edo (4,461,137), Enugu (4,396,098), Cross River (4,175,020), Abia (3,841,943), Ebonyi (3,007,155) and Bayelsa (2,394,725). The study

population includes all the males and female residents, including the skilled and unskilled manpower, employed, unemployed, self-employed and entrepreneurs in South East and South South Nigeria. But the sample size for the study was 385 gotten with the help of Australian Sample Size calculator (Australian Bureau of Statistics, 2024).

Face-to-face validity was used to validate the questionnaire used in this study. Two experts were consulted at Awka and Asaba to ensure for both face and content validity. Their advice was incorporated into the research instrument. To assess reliability, a test-retest method was employed, where the instrument was administered to 20 respondents from the zones and then re-administered after ten days. The obtained results were subjected to a reliability test using Pearson's Correlation Coefficient formula proposed by Francis Galton in 1880. The analysis convinced the researchers that the instrument was reliable because there was no significant difference between the answers supplied by the respondents at both the first and the second.

## 3.0. PRESEBNTATION OF RESULTS

Data extracted from the questionnaires were analysed using graphs and frequency of tables. Table 1 shows the respondents' demographic data as: 18-27 (21%); 28-37 (29%), 38-47 (24%) and 48 and above (26%). The respondents' gender was male (26%) and females (74%). Also, the data shows the respondents' State of residence as Abia (11%), Akwa Ibom (5%), Anambra (16%) and Bayelsa (9%). Others were Cross River (6%), Delta (15%), Ebonyi (5%), Edo (4%), Enugu (18%), Imo (10%) and Rivers (5%). Similarly, the respondents' education background was as follow: none (14%), First School Leaving Certificate (FSLC)-19%, National Diploma/National Certificate on Education (NCE-15%). Others were first degree or Bachelor degree (24%), Master's degree (17%), Doctorates Degree (PhD.) 11%. The respondents' occupation includes; student (32%), farmer (20%), trader (23%), Technicians/Artisans (6%) and public servant (25%).

**Table 1: Respondents Demographic Data**

| Responses | Frequencies | Percentages |
|---|---|---|
| **Age** | | |
| 18-27 | 83 | 21% |
| 28-37 | 111 | 29% |
| 38-47 | 92 | 24% |
| 48 and above | 99 | 26% |
| **Gender** | | |
| Male | 102 | 26% |
| Female | 283 | 74% |
| **States of Residence** | | |
| Abia | 44 | 11% |
| Akwa Ibom | 19 | 5% |
| Anambra | 56 | 15% |
| Bayelsa | 35 | 9% |
| Cross River | 23 | 6% |
| Delta | 52 | 14% |
| Ebonyi | 19 | 5% |
| Edo | 15 | 4% |
| Enugu | 63 | 16% |
| Imo | 38 | 10% |
| Rivers | 21 | 5% |
| **Education** | | |
| None | 53 | 14% |
| FSLC | 72 | 19% |
| Diploma/NCE | 59 | 15% |
| Bachelor degree | 94 | 24% |
| Master's degree | 64 | 17% |
| PhD. | 43 | 11% |
| **Occupation** | | |
| Student | 113 | 29% |
| Farmer | 71 | 18% |
| Trader | 83 | 22% |
| Technicians/Artisans | 23 | 6% |
| Public Servant | 95 | 25% |

Source: Authors' Fieldwork (2025).

Figure 1 shows the respondents knowledge of digital banking. According to the figure, 231 respondents (60%) are familiar with the concept of Digital Banking services (Internet banking, mobile banking) whereas 154 respondents (40%) are not aware. The respondents' also rated their awareness level of digital banking thus: Excellent (69 or 18%), Good (138 or 36%), Fair (102 or 26%) and Poor (76 or 20%). The data reveal the digital banking services respondents are more familiar with as Mobile banking app (83 or 22%), Internet/Online banks (71 or 18%), USSD banking (98 or 25%) and Fintech/online banks (133 or 35%).

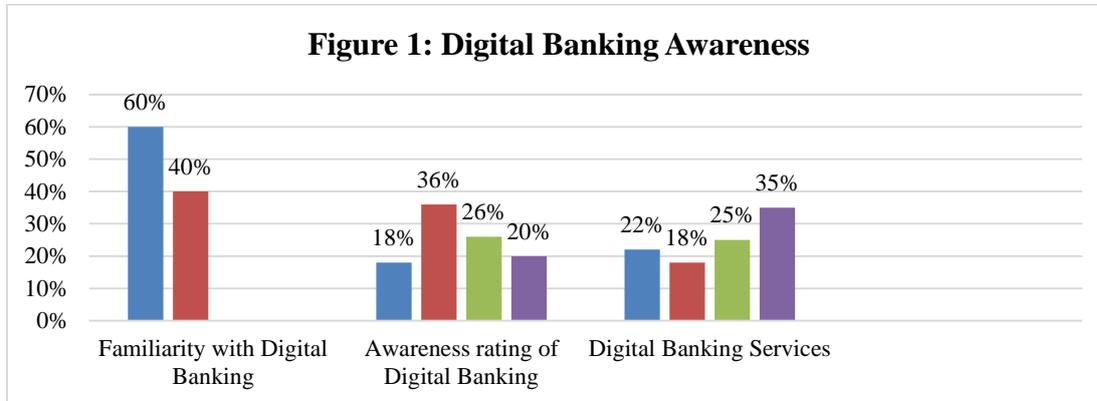

Figure 2 shows the respondents digital banking usage, frequency of its usage, reasons for usage and disadvantages. The data shows that the respondents using digital banking services were (313 or 81%) while 72 or 19% do not use it for their financial transactions. Also, the digital banking services respondents use most frequently were Commercial banks mobile banking app (64 or 20%), Commercial banks Internet banking platform (45 or 12%), USSD banking (61 or 22%) and Fintech/online banks (143 or 46%). The respondents listed their reasons for using digital banking most as: Convenience to access my account (51 or 16%), Faster for performing transactions (43 or 14%), Lower charges compared to traditional banking (62 or 20%), Improved security features (55 or 17%), and Easy access to loans (102 or 33%).

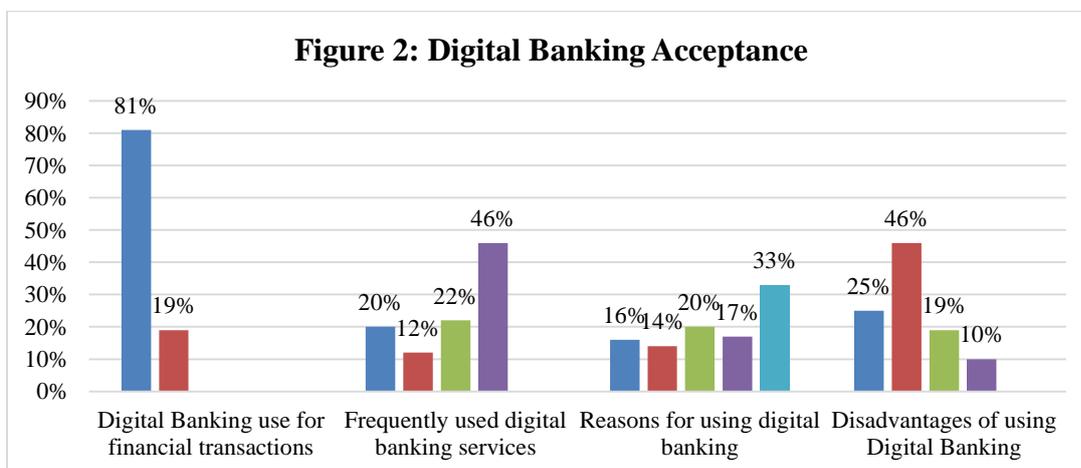

Source: Authors' Fieldwork (2025).

The disadvantages of using digital banking were enumerated as lack of access to smartphones (77 or 25%), hackers/Internet security concerns (144 or 46%), limited digital literacy/technical skills (61 or 19%), and unfamiliarity with the services (31 or 10%).

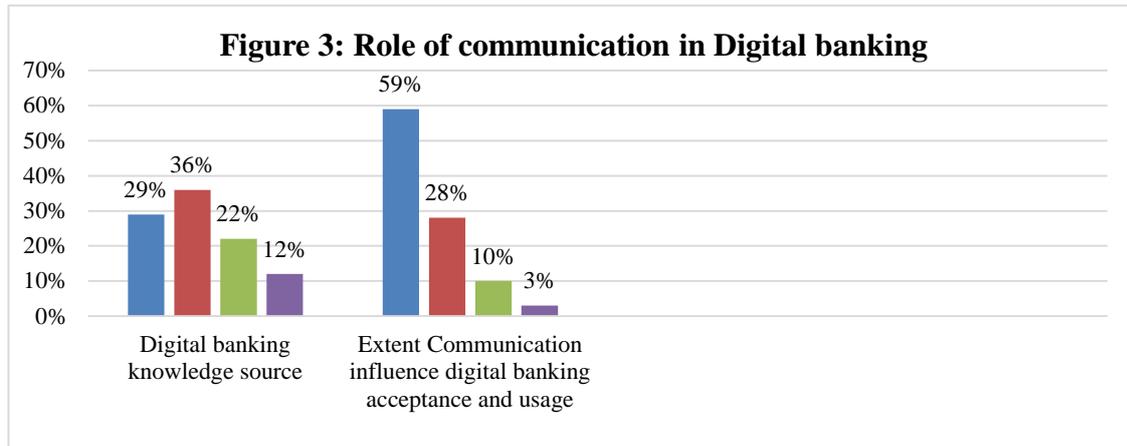

Source: Authors' Fieldwork (2025).

Figure 3 identifies the role of communication in digital banking. It shows that the respondents get to know about digital banking as banks customers care services (92 or 29%), media reports and advertisements (112 or 36%), family and friends (68 or 22%), and social groups/Institutions (41 or 12%). The respondents also disclose the extent communication influences their acceptance and usage of digital banking during the cash crunch as follows: to a great extent (186 or 59%), somewhat (87 or 28%), very little (32 or 10%) and not at all (8 or 3%).

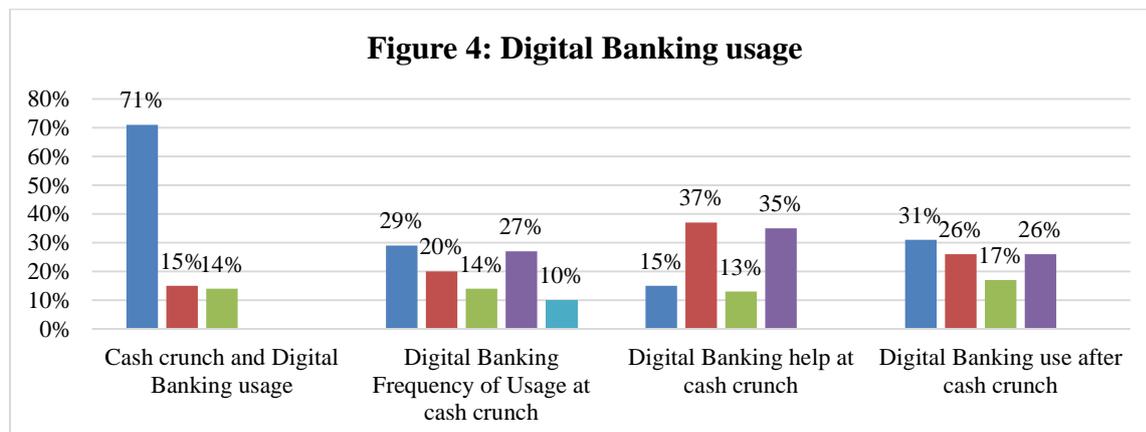

Source: Authors' Fieldwork (2025).

Figure 4 shows the respondents digital banking usage during and after the cash crunch in Nigeria. It shows that 221 respondents representing 71% used it during the recent cash crunch, 47 (15%) did not and 45 (14%) were unsure (14%) if they had used it. Also, the respondents frequency of using digital banking during the recent cash crunch were Daily (65 or 29%), Severally times a day (44 or 20%), Weekly (31 or 14%), Several times a week (59 or 27%) and Never (22 or 10%). Also, digital banking helped the respondents during the cash crunch in the following ways: accessing of account balance and transaction history (34 or 15%), bills and utilities settlements (82 or 37%), monetary transfers to others (28 or 13%) and cash withdrawal at ATMs/PoS (77 or 35%). The data also show the frequency respondents were using digital banking since end of cash crunch in Nigeria as daily (69 or 31%), severally times a day (57 or 26%), weekly (37 or 17%) and several times a week (58 or 26%).

## 4.0. DISCUSSION

The study was focused on Communication, Awareness and Acceptance of Digital Banking amidst Cash Crunch in South East and South-South, Nigeria. The findings (in figure 1) showed that awareness level of South-East and South-South Nigeria are good (138 or 36%). The data also revealed that the digital banking services respondents from the zones are more familiar with is Fintech/online banks (133 or 35%), followed by USSD banking (98 or 25%) and Mobile banking app (83 or 22%). This entails why the Central Bank of Nigeria (CBN) licensed over 894 microfinance Bank with many operating digital banking (Akintaro, 2023). A study by Neves, Oliveira, Santini and Gutman (2023) also confirmed that adoption and usage of different digital financial services were on the rise across the globe with India emerging on top (22%), followed by Brazil (6%) and China (5%). The implication is that many Nigerians will in future depend solely on digital banking services over conventional banking services due to cashless policies and other reasons.

It was also determined that acceptance level of digital banking among the South-East and South-South Nigeria was high as 313 or 81% admitted accessing it in Figure 2. The respondents also listed the digital banking services they frequently accessed as Fintech/online banks (143 or 46%), followed by USSD banking (61 or 22%) and commercial banks mobile banking app (64 or 20%). In contrary, Wezel and Ree (2023) findings showed that although Nigeria was making progress in digital financial services the rate in which the populace embraces it was poor due to low financial literacy level. This implied that the media and communication were needed to enlightening and boosts the masses with digital financial knowledge so as to utilise it well.

The role of communication in awareness and acceptance of digital banking during the cash crunch in South-East and South-South Nigeria was found in figure 3. The data showed that communication helps the respondents to access information for digital banking mostly through media reports and advertisements (112 or 36%), followed by banks customers care services (92 or 29%) and family and friends (68 or 22%). The extent communication assists the respondents to access information about digital banking is to a great extent (186 or 59%). This finding was not different from Kitsios et al. (2021) discovery that technology promotes ease of use, visibility and acceptance of digital banking. This implies that most people using digital banking services in the South-East and South-South Nigeria are doing that because of communication which makes it easier for their understanding and application of it. It also called for more enlightening campaign and media sensitisation to improve the usage of the financial innovation.

The study also confirmed that there was usage of digital banking services during the cash crunch but it became more at the end of the financial crises. According to the data in figure 4, (221 or 71% respondents) used it during the recent cash crunch. Also, the respondents frequency of using digital banking during the recent cash crunch were several times a week (59 or 27%) and daily (69 or 31%) after the cash crunch. The findings in this study is inline with Indriasari, Prabowo and Purwandari (2022) where it was found that acceptance and usage of digital banking improved during and after COVID-19. They also found that mobile devices and Internet contributes to increase in usage of digital banking and other financial services. Also like cash crunch, Kitsios et al. (2021) also found that use of Digital Banking became more prominent in COVID-19 pandemic. This implied that digital banking usage could help to overcome emergency situations and crisis such as disease epidemic, among others.

## 5.0. CONCLUSION AND RECOMMENDATIONS

### 5.1. Conclusion

This study was aimed at ascertaining the role of communication in determining awareness and acceptance of digital banking amidst cash crunch in South East and South-South, Nigeria. However, it was limited by insufficient data on digital banking and communication, especially in the zones studied. Other limitations to the study are time and finances which prevented the researchers from exploring other methodologies like focus group discussion and interview. It was, however, found that digital banking services were used during the cash crunch and the usage had been on increase since end of the cash crunch in South-East and South-South, Nigeria. The study therefore concludes that usage of digital banking in South-East and South-South Nigeria are on increase.

### 5.2. Recommendations

The researchers hereby recommended increase sensitisation of the residents in the zones on importance of digital banking services using both the mainstream and the new media for increased acceptance and usage of it. Researchers are also advising others to explore other areas of digital banking like digital currency, cryptocurrency and so on.

### Competing Interest

The authors have declared that no conflicting interest exist in this paper.

## REFERENCES

Adeniyi, O. (2021, June 14). Why digital banking is the key to transforming Nigeria's f financial landscape. *Techcabal. https://techcabal.com/2021/06/14/why-digital-banking-is-the-key-to-transforming-nigerias-financial-landscape/*

Akintaro, S. (2023). Here are 10 digital banks licensed by the CBN to operate as microfinance banks in Nigeria. *Nairametrics*. https://www.google.com/amp/a/nairametrics.com/2023/03/22/here-are-10-digital-banks-licensed-by-the-cbn-to-operate-as-microfinance-banks-in-nigeria/%3famp=1

American Bankers Association. (2023, September 12). New ABA report: 88% of Banks report active social media engagement. *ABA banking journal.* https://bankingjournal.aba

.com/2023/09/aba-report-most-banks-report-active-social-media-engagement/

Australian Bureau of Statistics. (2024). Sample size calculator. https://www.abs.gov.au/websi
tedbs/d3310114.nsf/ home/sample%20size%20calculator

Das, R., S. (2022, December 27). A practical guide to social media for banks in 2024.
https://statusbrew.com/insights/social-media-for-banks/

Guardian newspaper. (2023). Digital banking in Nigeria is on the rise in 2023. *Guardian*.
https://m.guardian.ng/digital-banking-in-nigeria-is-on-the-rise-in-2023/

Indriasari, E, Prabowo, H. & Purwandari, B. (2022). Digital banking: challenges, emerging
technology trends, and future research agenda. *International journal* of *E-Business
research*, 18(1), 1-19. https://doi.org/10.4018/IJEBR.309398

Joseph, B. (2023, July 18). Nigeria's cash crunch sparks increase in digital payments,
creating opportunities for Merchant Acquirers. *Msmeafrica*. https://msmeafricaonline.c
om/nigerias-cash-crunch-sparks-increase-in-digital-payments-creating-opportunities-
for-merchant-acquirers/

Kazim, W. (2024, October 7). Digital transformation in Banking: Trends and opportunities
*G2 media*. https://www.g2.com/articles/digital-transformation-in-banking

Kitsioss, F., Giatsidis, I. & Kamariotou, M. (2021). Digital transformation and strategy in the
Banking sector: evaluating the acceptance rate of E-Services. *Journal of open
innovation*: *technology*, *market and complexity*, 7(3), 204. https://doi.org/10.3390/joitm
c7030204

Lottu, O. A., Abdul, A. A., Daraojimba, D. O., Alabi, A. M., John-Ladega, A. A. &
Daraojimba, C. (2023). Digital transformation in banking: a review of Nigeria's
journey to economic prosperity. *International* journal *of advanced* economics,
5(8), 215-238. https://doi.org/10.51594/ijae.v5i8.572

Napoletano, E. & Foreman, D. (2021, February 24). What is digital banking? *Forbes*.
https://www.forbes.com/councils/forbestechcouncil/2024/05/01/the-rise-of-digital
-banking-a-paradigm-shift-in-fintech/

Neves, C, Oliveira, T, Santini, F & Gutman, L. (2023). Adoption and use of digital financial
services: A meta analysis of barriers and facilitators. *International journal of
information management data insights*,3(2)-100201. https://doi.org/10.1016/j.jjimei.20
23.100201

Nigeria Bureau of Statistics. (2021). E-payment Channels: Quarterly Summary Q1 2021.

Nigerian static demographic bulletin. (2020). Nigerian demographic.
https://nigerianstat.gov.ng/pdfuploads/DE MOGRA PHIC%20BULLETIN%202020.pdf

Salawu, D., Oyebayo, D. & Nwachukwu, H. (2021, October 10). Digital banking in Nigeria: a dive into the State of play and legal framework. *Olaniwunajayi*. https://www.olaniwu najayi.net/blog/wp-content/uploads/2021/10/Digital-Banking-in-Nigeria-A-Dive-Into -the-State-of-Play-and-Legal-Framework-in-Nigeria-New.pdf

Wezel, T. & Ree, J. (2023, March 6). Nigeria fostering financial inclusion through digital financial services. *International Monetary Fund (IMF) country report 2023*. https://www.imf.org/en/Publications/selected-issues-papers/Issues/2023/03/07/ Nigeria-Fostering-Financial-Inclusion-through-Digital-Financial-Services-Nigeria- Nigeria-530633